\newcommand{\vct}[1]{{\bf #1}}
\newcommand{\eq}[1]{Eq.~(\ref{eq:#1})}
\newcommand{\fig}[1]{Fig.~(\ref{Fig:#1})}

\newcommand{\tab}[1]{Table~\ref{tab:#1}}
\newcommand{\olc}[1]{Ref.~\onlinecite{#1}}

\documentclass[preprint,aps,prl,showpacs,preprintnumbers,amsmath,amssymb]{revtex4-1}

\usepackage{graphicx}
\usepackage{dcolumn}
\usepackage{bm}
\usepackage[usenames]{color}
\usepackage{slantsc}
\usepackage{dsfont}
\newcolumntype{d}{D{.}{.}{2.2}}
\usepackage{siunitx}

\begin{document}


\title{Elastic constants and supersolidity in solid $hcp$ $^4$He}

\author{Renato Pessoa}
\affiliation{Instituto de F\'{\i}sica, Caixa Postal 131 \\
Universidade Federal de Goi\'as (UFG) \\
74001-970, Goi\^ania, GO, Brazil}
\author{M. de Koning}
\author{S. A. Vitiello}
\affiliation{Instituto de F\'{\i}sica Gleb Wataghin, Caixa Postal 6165 \\
Universidade Estadual de Campinas - UNICAMP \\
13083-970, Campinas, SP, Brazil}

\date{\today}

\begin{abstract}
The elastic constants of solid \textit{hcp} $^4$He are investigated in molar volumes ranging from about melting up to approximately 14 MPa. Properties of interest are determined by averages formed from computed values in configurations sampled by Monte Carlo of a model wave function. Deviations from known elastic relations are reported near the density where the supersolid fraction is maximum. The results offer further evidence that the supersolid state is related to elastic constants anomalies and that both are manifestations of a single physical process.
\end{abstract}


\maketitle


For more than a hundred years, since it has been liquefied, helium has shown intriguing behaviors and has provided invaluable clues for the condensed matter physics. Nowadays, despite a reasonable good understanding of its liquid phase, solid helium is still revealing fascinating aspects that defy experimentalists and theoreticians. The shear modulus softening with temperature increase, as observed by Day and Beamish\cite{day07}, and its connections to supersolidity, superfluidity superimposed to crystalline order, as seen in the pioneering experiments by Kim and Chan\cite{ kim04,kim04b} illustrate very well this situation.

Elasticity is ubiquitous among materials and has been extensively investigated in solids where atomic exchange is negligible, and the atoms are reasonable localized and distinguishable. If forces that are not too strong are applied in the surfaces of these solids, they will produce mechanical deformations that will increase the potential energy. As the solid is released from these forces and assuming they are small enough, it will go back to its original state. However, what can be said about quantum solids? Systems where the zero-point kinetic energy is comparable to the binding energy of the atoms that perform large excursions from their mean positions in the lattice. How will these quantum systems respond to applied forces?

This letter reports values of all the five independent elastic constants, as a function of the density, which characterize a defect-free single crystal of $^4$He in an \textsl{hcp} structure. We have analyzed our results using relations among the elastic constants that are satisfied under well defined conditions. Since all the elastic constants were determined, it was possible to some extent verify the internal consistence of our calculations and search for elastic anomalies in this quantum solid. In fact, we have observed deviations of the relations occurring near the same density where the mass decoupling in the movement of torsional oscillators has its maximum.\cite{kim06} In our view these results are further evidences  that supesolidity\cite{kim04,kim04b} and elastic anomalies\cite{day10,day07} of the shear modulus are related phenomena and consequence of a single physical process.

The elastic constants were calculated at zero temperature using a model wave function that has allowed an excellent description of various properties of $^4$He in both the liquid and solid phases.\cite{caz12,ros09} It is worth mentioning that the interacting potential for the helium atoms is very accurately known as more than one study has shown.\cite{uje05,uje06} These facts, together with Monte Carlo calculations performed with a reasonable number of atoms, assure that the estimated quantities can have good reliability.


A system formed from $^4$He atoms of mass $m$ can be well described by the Hamiltonian
\begin{equation}
 H = -\frac{\hbar^2}{2m}\sum_i^N\nabla_i^2 + \sum_{i<j}V(r_{ij})\;,
\label{eq:hamiltonian}
\end{equation}
where $V(r)$ is the Hartree-Fock dispersion two-body (HFD-HE2) potential of Aziz and collaborators.~\cite{azi79}  This inter-atomic interaction is still widely used,\cite{del11,pes10} even if more accurate potentials exist.\cite{hur07,jan97} The HFD-HE2 potential was kept in our calculations because it does not seem that our results will be significantly modified if the more recent potentials were considered. Moreover there is already a large body of work performed with the model function used in this work that relays on the HFD-HE2 inter-atomic potential.

We characterize the solid phase by a shadow wave function\cite{vit88,mac94} (SWF) written as
\begin{widetext}
\begin{equation}
 \Psi (R)= \prod_{i<j}^N \exp \left [ - \frac{1}{2} \left( \frac{b}{r_{ij}}\right )^5\right ] \int dS \prod_{i}^N\exp \left [ -C \vert {\bf r}_i - {\bf s}_i\vert^2 \right ] \prod_{i<j}^N \exp \left [ - \delta V(\alpha s_{ij})\right]
\label{eq:swf}
\end{equation}
\end{widetext}
where $R\equiv \left \{  {\bf r}_1, {\bf r}_2 ,\ldots,{\bf r}_N\right \}$,
$S\equiv \left \{  {\bf s}_1, {\bf s}_2 ,\ldots,{\bf s}_N\right \}$,
and $dS \equiv \left \{ d{\bf s}_1 ,d{\bf s}_2, \ldots, d{\bf s}_N\right \}$. The integrations over the auxiliary variables, also called shadow particles, are performed in the whole space. The parameters $b$, $C$, $\delta$ and $\alpha$ are chosen so that the expectation value of the energy is minimized. The correlations between the shadow particles are imposed through the HFD-HE2 potential rescaled in its amplitude and inter-particle distance through the parameters $\delta$ and $\alpha$, respectively. This model wave function is considered one of the best available descriptions for systems consisting of helium atoms.\cite{caz12,ros09} It has been used to study several helium properties\cite{pes10,pes09br,ros08} including the structure and mobility of linear defects in a calculation of the Peierls-Nabarro stress.

Elastic constants are computed using the Parrinello and Rahman method.\cite{par80,par81} This treatment allows periodic boundary conditions to be considered  together with distortions that must be applied in the simulation cell in a easy way. If the orthogonal vectors $\mathbf{L}_x$, $\mathbf{L}_y$ and $\mathbf{L}_z$ define this box along the directions $(x,y,z)$, a position $\vct r_i$ in the cell can be written as
\begin{equation}
\mathbf{r}_i = \xi_i \mathbf{L}_x + \eta_i \mathbf{L}_y + \zeta_i \mathbf{L}_z = \mathds{h}_0 {\mathbf{s} }_i \;\;,
\label{r0}
\end{equation}
where the coefficients $\xi_i$, $\eta_i$ and $\zeta_i$ range between $0$ and $1$ and can be considered components of a vector ${\mathbf{s} }_i$. Then the vector $\mathbf{r}_i$ can be written as a product of a matrix $\mathds{h}_0$ by $\mathbf{s}_i$. The matrix $\mathds{h}_0$ is diagonal with elements equal to the modulus of the vectors that define the box.

Homogeneous deformations change the positions $\vct r_i$. In our calculations the new positions are obtained through a non-diagonal matrix $\mathds{h}$
\begin{equation}	
 \textbf{r}'_i = \mathds{h}{\bf \mathsf{s} }_i = \mathds{h}\mathds{h}^{-1}_0 \mathbf{r}_i \;.
\end{equation}
The strain-tensor then can be written as
\begin{equation}
 \epsilon_{\alpha \beta} = \frac{1}{2} \left ( {\mathds{h}^{\dag}_0}^{-1} \mathcal{G} \mathds{h}^{-1}_0 -  \mathbb{I}\right),
\label{strain}
\end{equation}
where the Greek symbols stand for any of the coordinate axes, $\mathcal{G}\equiv \mathds{h}^{\dag} \mathds{h}$ denotes a metric tensor and $\mathbb{I}$ is the unit matrix.

The stress-tensor $ \sigma_{\alpha\beta}$ is calculated through the virial theorem.\cite{cep77} It is estimated by averages of values calculated from configurations sampled from the square of the SWF using the Metropolis algorithm.  In the linear elastic regime, the stress and strain tensors are related by the generalized Hooke's law
\begin{equation}
\sigma_{\alpha\beta} = \sum_{\gamma,\chi} c_{\alpha\beta\gamma\chi} \epsilon_{\gamma\chi} ,
\label{eq:hooke}
\end{equation}
where the $c_{\alpha\beta\gamma\chi}$ are the stress-strain coefficients. The elastic constants,
\begin{equation}
C_{\alpha\beta\gamma\chi} = \frac{d \sigma_{\alpha\beta}}
{d \epsilon_{\gamma\chi}},
\end{equation}
were obtained by computing the slopes of stresses as a function of strains for selected deformations that have a single  strain-stress coefficient different from zero. Elastic constants are equal to the stress-strain coefficients if no hydrostatic pressure $P$ is applied to the crystal. Otherwise the elastic constants are related to stress-strain coefficients by\cite{sti01}
\begin{equation}
C_{\alpha\beta\gamma\chi} =  c_{\alpha\beta\gamma\chi} + P(
\delta_{\beta \alpha} \delta_{\alpha \gamma} 
+\delta_{\alpha \chi} \delta_{\beta \gamma} 
-\delta_{\alpha \beta} \delta_{\gamma \chi} 
).
\label{eq:celas} 
\end{equation}

Relations among the elastic constants were used\cite{fra70} in the past to complete their individual determination from experimental data. Here, since we are able to calculate all of these constants independently one from the others and with good accuracy, the relations are used to validate the internal consistence of our calculations and to look for deviations that might signalize elastic anomalies.

The generalization of Hooke's law considered together with the symmetric character of the stress tensor, the reversibility of the deformation work and the symmetry of a \textsl{hcp} crystal result in a system with only five independent elastic constants. If the additional constraint of a short-range central interacting potential is assumed, the number of elastic constants is reduced to four independent values. This additional symmetry is one of the Cauchy relations that is still valid in an \textsl{hcp} structure\cite{bor54}

\begin{equation}
c_{13} = c_{44}.
\label{eq:cauchy} 
\end{equation}
From now on we use the Voigt notation.\cite{nye85} It is worthwhile to note, as it has been extensively discussed in the literature, that the Cauchy relations are not rigorously fulfilled by any material. Nevertheless this fact has not prevented its use as a source of informations, for instance, about bonding properties  of the system.\cite{zha04,fra70}

Another relation among elastic constants can be obtained by considering the linear compressibility. It is given by the relative longitudinal change in length of a given direction with respect to the hexagonal axis when the system is under unit hydrostatic pressure. In general this quantity is non-isotropic. However if the ratio $c/a$ of the lattice parameters along the hexagonal axis $c$ and the basal plane $a$ is independent of the pressure, the linear compressibility is isotropic and as a consequence the relation

\begin{equation}
c_{11} + c_{12} = c_{33} +c_{13}
\label{eq:lc} 
\end{equation}
is satisfied.


We have computed properties of interest at several molar volumes  resulting in densities varying between 0.0292 and 0.0353~\AA$^{-3}$. Simulation cells we have used contained 720 atoms in an ideal \textit{hcp} structure.  The equilibration process consisted of $2\times 10^4$ Monte Carlo (MC) sweeps (attempts to move randomly all the particles and shadows coordinates). At all densities, the parameters of the SWF are those that minimize the expectation value of the Hamiltonian operator of \eq{hamiltonian}. The variational search in the parameter space was performed either by us or is reported in the literature.\cite{mac94}

The stress tensor components were estimated by performing simulations of at least $1.6\times10^6$ MC sweeps for each of the different strains. Each elastic constant was determined using four different distortions in the simulation cell. The maximum strain levels we have considered was $0.25\%$. 

\begin{figure}
\includegraphics*[angle=-90,width=1.0\linewidth]{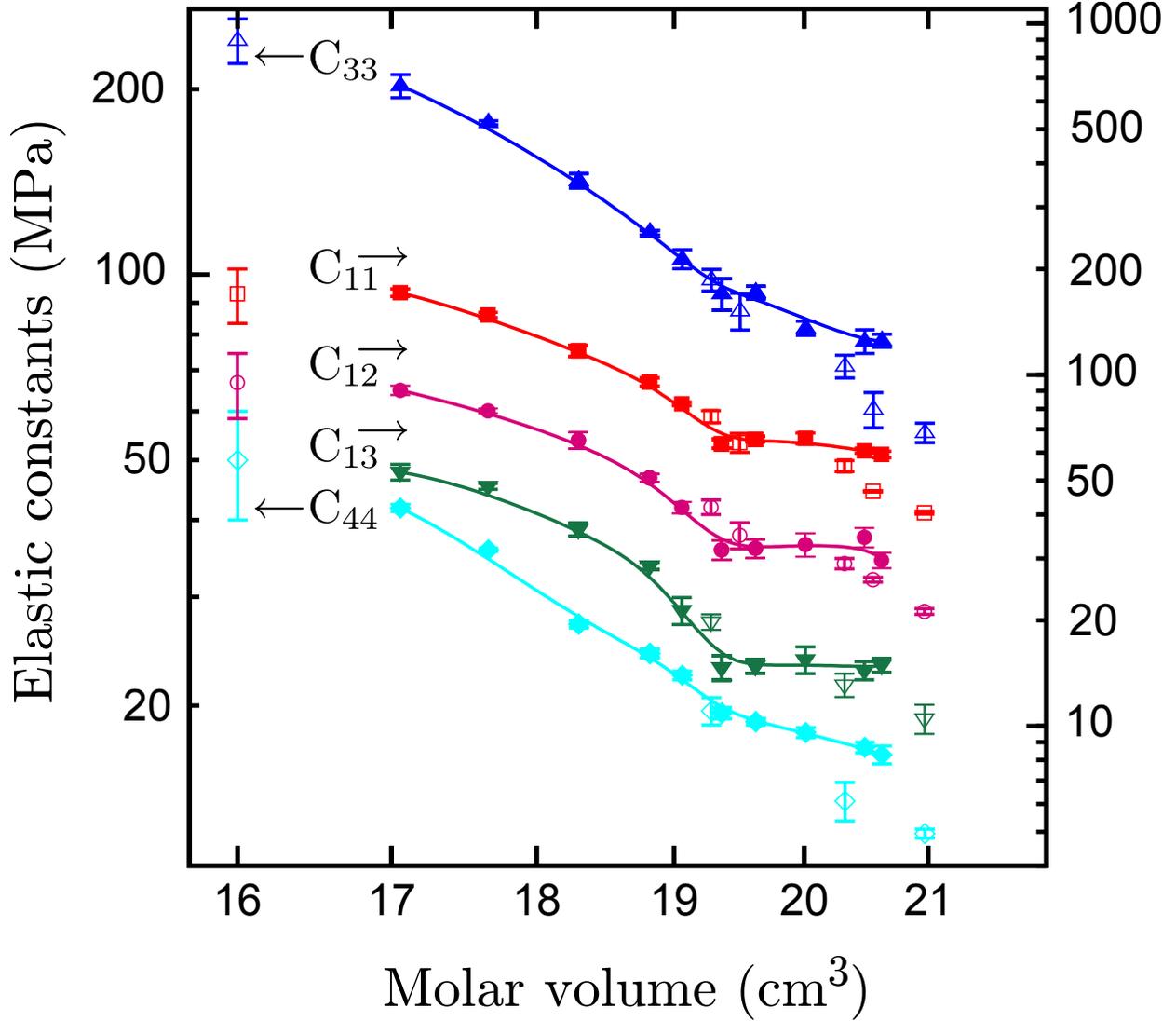}
\caption{\label{Fig:constants} Elastic constants of $^4$He (solid symbols) as a function of the molar volume in a log-log plot. The empty symbols stand for the experimental data \cite{cre71,gre71,wan70,fra70,ree71,bea01}. Values are from the left or right hand scales according to the arrows in the figures. The lines are guide to the eyes.}
\end{figure}

The elastic constants were determined in molar volumes ranging from 20.66 cm$^3$, near the melting density, to 17.06 cm$^3$. The results are displayed in \fig{constants} together with the data from the literature.\cite{cre71,gre71,wan70,fra70,ree71,bea01} The general agreement between theory and the few experimental data points is good. From the figure we can see that constants $C_{11},\ C_{12}$ and $C_{13}$ have variations with respect to density that clearly show some sort of discontinuity near the molar volume 19.36 cm$^3$. Although constants $C_{33}$ and $C_{44}$ seem also to show this feature at the same density, it is much less pronounced. It is worthwhile mentioning that near this density, a maximum in the supersolid fraction\cite{kim06} has been observed.

After completing this work we became aware of \olc{caz12} where elastic constants were investigated by diffusion Monte Carlo.  In this paper the authors point out that the elastic constants that we have already published\cite{pes10} at a single density $\rho = 0.0294$ \AA$^{-3}$ is in good agreement with their results. At other densities we see a less satisfactory agreement. However the most important disagreement  between our results are the overall behavior of the elastic constants with pressure. While in \olc{caz12} the elastic constants increase linearly with pressure, our results are far from this behavior. Especially for the elastic constants $C_{11}$, $C_{12}$ and $C_{13}$, we see roughly constants values up to $\rho = 0.0311$ \AA$^{-3}$ where the magnitudes start to increase with a character that is hardly linear. We do not know the reasons for such discrepancies. However it would be difficult to attribute them to our model. It has been recognized as excellent.\cite{caz12} Eventually they might be due to finite-size effect. Although the authors of \olc{caz12} have made a careful analises of this effect on the energy, it is well known that simulations made to obtain the elastic behavior of materials require large systems. Our system is more than three times bigger than theirs.

In \tab{elastic_constants} at each one of the given densities, we report values determined in our calculations for the elastic constants, the total energies per atom and the pressures, obtained by taking the trace of the stress-tensor.

\begin{table*}
\resizebox{0.95\linewidth}{!}{%
\begin{tabular}{ccccccccc}
\hline
 $\rho~(\textrm{\AA}^{-3})$&  $V$ (cm$^3$)& E (K) & $P$ (MPa)&$C_{11}$ & $C_{33}$ & $C_{44}$ & $C_{12}$ & $C_{13}$ \\
\hline
 $0.0292$ & $20.62$ & $-5.0770 \pm 0.0010$ & $ 2.4529 \pm 0.0047$ & $59.3\pm1.2\;\;$ & $78.1\pm1.9\;\;$ & $16.63\pm0.56\;\;$ & $29.6\pm1.5\;\;$ & $14.89 \pm 0.68$\\ 
 $0.0294$\footnote[2]{Results from Ref.~\onlinecite{pes10}} & $20.48$ & $-5.0349 \pm 0.0007$ & $3.0957 \pm 0.0026$ & $60.8\pm1.0\;\;$ & $77.9\pm3.4\;\;$ & $17.10\pm0.34\;\;$ & $34.4\pm2.2\;\;$ & $14.40\pm0.85   $ \\
 $0.0301$ & $20.01$ & $-4.8646\pm 0.0007$ & $4.0707 \pm 0.0028$ &$65.7\pm1.8\;\;$ & $82.9\pm2.3\;\;$ & $18.07\pm0.34\;\;$ & $32.9\pm3.1\;\;$ & $15.8\pm1.3$ \\
 $0.0307$ & $19.62$ & $-4.6410 \pm 0.0007$ & $5.0941 \pm 0.0029$ & $65.5\pm1.2\;\;$ & $93.5\pm2.2\;\;$ & $18.83\pm0.22\;\;$ & $32.1\pm1.9\;\;$ & $14.81\pm0.67$ \\
 $0.0311$ & $19.36$ & $-4.4799 \pm 0.0008$ & $5.7905 \pm 0.0031$ & $63.6\pm1.8\;\;$ & $93.1\pm5.5\;\;$ & $19.43\pm0.39\;\;$ & $31.7\pm2.0\;\;$ & $14.7\pm1.1$ \\
 $0.0316$ & $19.06$ & $-4.2864 \pm 0.0007$ & $6.5933 \pm 0.0030$ &$82.7\pm1.0\;\;$ & $105.9\pm3.7\;\;$ & $22.38\pm0.36\;\;$ & $41.9\pm1.6\;\;$ & $21.3\pm1.9$ \\
 $0.0320$ & $18.82$ & $-4.0713 \pm 0.0007$ & $7.2055 \pm 0.0031$ & $95.4\pm2.6\;\;$ & $116.9\pm1.0\;\;$ & $24.29\pm0.39\;\;$ & $50.9\pm1.2\;\;$ & $28.61\pm0.68$ \\
 $0.0329$ & $18.30$ & $-3.5904 \pm 0.0017$ & $8.8184 \pm 0.0081$ & $117.0\pm4.2\;\;$ & $142.1\pm3.7\;\;$ & $27.10\pm0.35\;\;$ & $65.1\pm3.4\;\;$ & $36.3\pm1.5$ \\
 $0.0341$ & $17.66$ & $-2.8139 \pm 0.0006$ & $11.6933 \pm 0.0035$ &$148.0\pm2.5\;\;$ & $176.3\pm1.3\;\;$ & $35.77\pm0.22\;\;$ & $78.9\pm1.2\;\;$ & $48.16\pm0.95$ \\
 $0.0353$ & $17.06$ & $-1.9265 \pm 0.0013$ & $14.1040 \pm 0.0074$ & $171.3\pm4.3\;\;$ & $202.4\pm8.7\;\;$ & $41.82\pm0.57\;\;$ & $90.4\pm2.7\;\;$ & $52.8\pm2.7$ \\
\hline             
\end{tabular}
}      
\caption{Elastic constants $C_{ij}$ in MPa calculated at the given densities (first column) in solid $hcp$ $^4$He. The second column shows molar volumes. The third and fourth columns show the values of energies per atom and pressures.}
\label{tab:elastic_constants}
\end{table*}

\begin{figure}
\includegraphics*[angle=-90,width=0.95\linewidth]{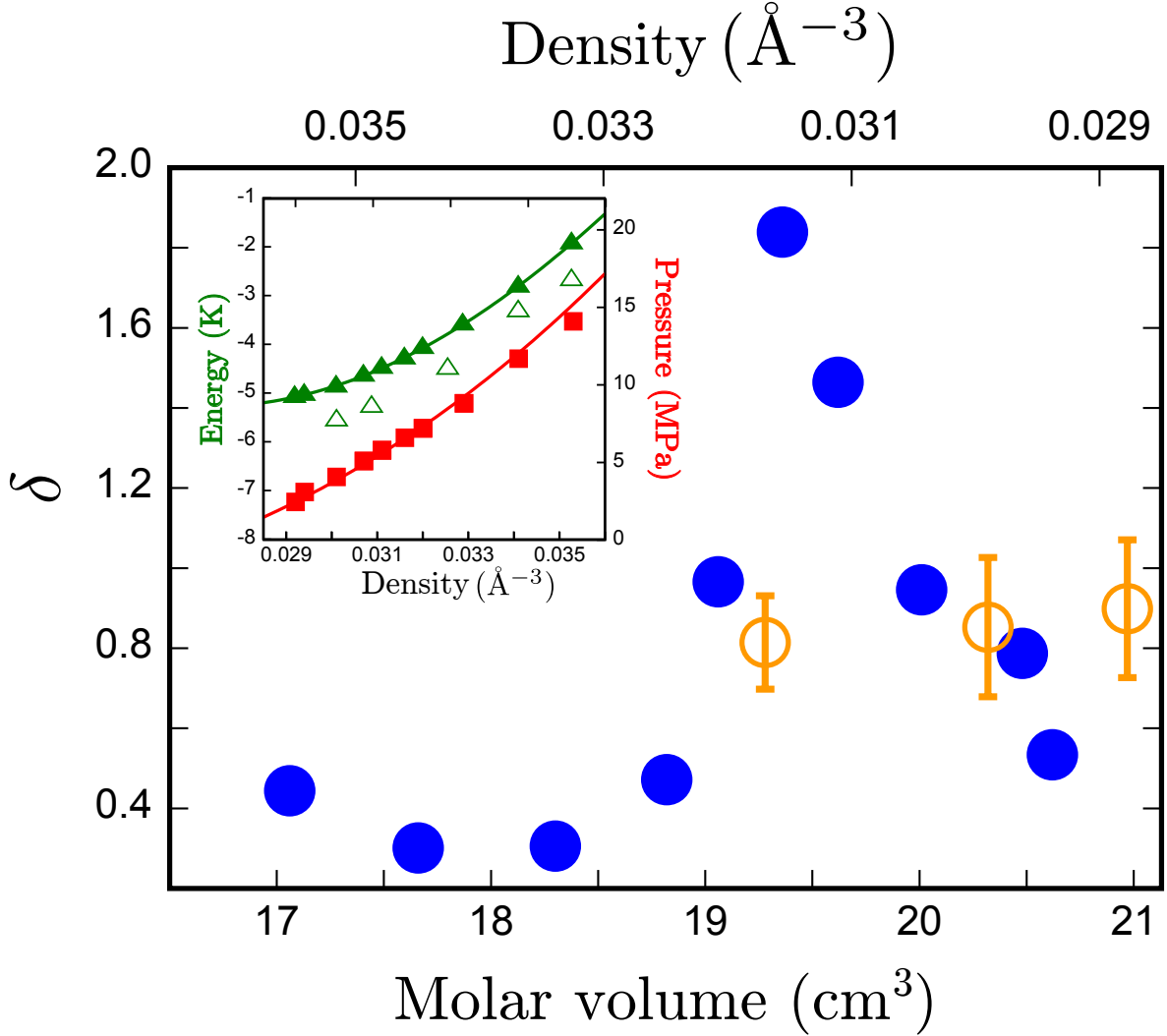}
\caption{\label{Fig:delta} The ratio $\delta$ that assess deviations of the Cauchy relation as a function of the molar volume and density, respectively lower and upper scales. The empty circles stand for the experimental data~\cite{bea01}. The inset shows the computed energies per atom as a function of the density (solid triangles and left-hand side scale) and the experimental data \cite{edw65} (open symbols). The line is a fourth order polynomial fit to the energies. The pressure (solid squares) has its units in the right-hand side. The line is obtained through the derivative of the EOS, $P(\rho) = \rho^2 \partial(E) / \partial \rho$}
\end{figure}

We have analyzed our results considering the relations among the elastic constants already discussed. For this purpose it is useful to rewrite \eq{cauchy} using \eq{celas} and define

\begin{equation}
\label{eq:delta}
\delta = \frac{(C_{44} + P) - (C_{13} - P)}{(C_{13} - P)}\;,
\end{equation}
this is a quantity that depends on the shear modulus $\mu = C_{44}$. In principle $\delta$ is zero if  zero-point motion effects and many-body components in the inter-atomic potential were negligible.  As already point out, the general validity of the Cauchy relation that originates \eq{delta} might be questionable. However whatever value $\delta$ could have it would be hard to expect abrupt changes in its magnitude as the pressure is slightly varied.  \fig{delta} presents the values we have  computed for $\delta$. First of all, let us note that at both ends of the densities range we have considered, $\delta$ is approximately 0.4. A possible value, because in a system made from $^4$He atoms the requirements for the validity of the Cauchy relation, $\delta=0$, would hardly be met. However near the molar volume 19.36 cm$^3$, $\delta$ has a peak with a maximum value equal to  almost 2.0. This is near the density where the supersolid fraction has a maximum value.\cite{kim06} The quantity $\delta$ has also been computed using the available experimental data.  Values obtained from experiment are in good agreement with theory. However there are experimental data only at few densities, what precludes a detailed comparison between theory and experiment at the density where $\delta$ show its peak.

The behavior of $\delta$ can not be imputed to an artifact of our model. As a possible mean to verify this fact we have considered the equation of state (EOS) of the total energy per atom as a function of the density. The computed EOS is in very good agreement with experimental data as it is possible to verify in the inset of \fig{delta}. The computed pressures are within the values one expects for this quantity. It is  also displayed in the inset, and it does not show neither any signal pointing towards an specific difficulty of the model.

\begin{figure}
\includegraphics*[angle=-90,width=0.95\linewidth]{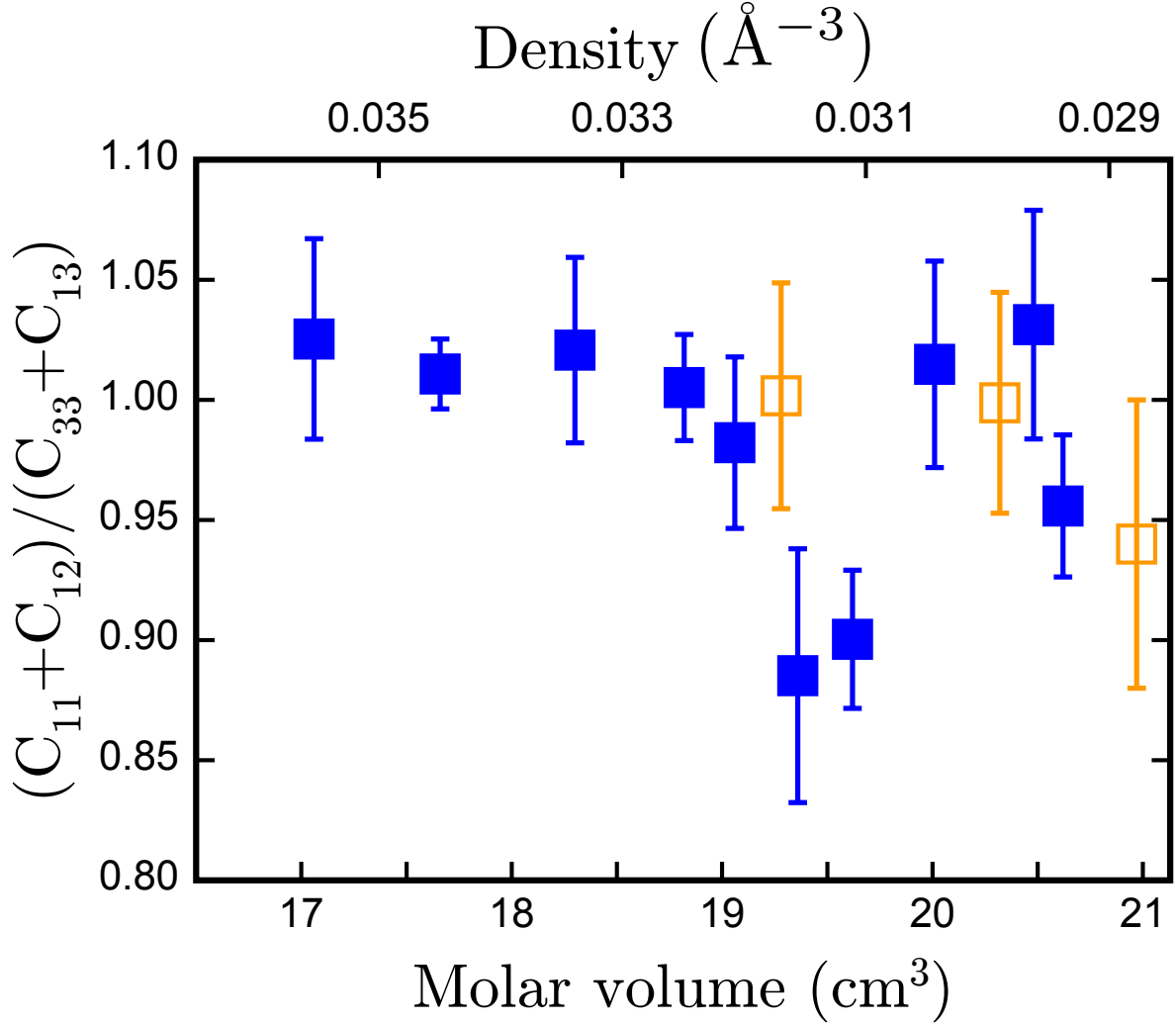}
\caption{\label{Fig:ratio} The elastic constants ratio $(C_{11} + C_{12})/(C_{33}+C_{13})$ as a function of the molar volume at the lower scale and the density in the upper scale. The solid symbols stand for our theoretical results and those empty for the experimental data.\cite{bea01}
}
\end{figure}

For solid $^4$He as we can learn from experiment\cite{lou93} and theory,\cite{caz12} the axis ratio $c/a$ is almost pressure independent and very near 1.63, the value $c/a$ has in an ideal \textsl{hcp} crystal. In this context, we expect the relation expressed in \eq{lc} to hold. Because of \eq{celas}, this relation is valid for both the stress-strain coefficients and the elastic constants. We have computed the ratio ${\cal R} =(C_{11} +C_{12})/(C_{33}+C_{13})$, as a way of verifying the validity of the relation of \eq{lc}  and plotted our results in \fig{ratio}. Since $^4$He has the ratio $c/a$ practically constant and independently of the pressure, we expect the ratio $\cal R$ to have also a constant value equal to one. In fact we see this behavior at most of the densities where the calculations were performed. However, as we see in the figure, $\cal R$ deviates from this value.  This quantity has a minimum in a form that resembles a peak upside down at the same density where the supesolid fraction and $\delta$ of \eq{delta} have their maxima. We also have computed the ratio $\cal R$ using experimental data and once again only in a few densities there are available data. This is not a helpful situation for a direct comparison between experiment and theory where we see the largest deviation of $\cal R$. However where the comparison is possible the agreement is good.

In this work, as we have described a \textsl{hcp} defect-free single crystal of $^4$He, the elastic constants were computed as a function of the density.  The deviations from the relation of  \eq{lc}, derived from the well established fact that $^4$He has an isotropic linear compressibility, show according our view that our calculations are consistent with experimental observations that a system formed from these atoms present elastic anomalies. At the same time this relation being obeyed at the lowest and highest densities we have considered offer a confirmation of the consistence of our calculations. The behavior $\delta$ defined in \eq{delta} corroborates this fact. The values of this quantity are able to display the zero-point motion effects at the extremes of the density range and an elastic anomaly, as exhibited by a relatively large peak. We do not believe that values of $\delta \approx 0.4$  could be attributed to a non-central character of the atomic interaction. Atomic three-body interactions  can be accurately computed\cite{uje03} and we know\cite{uje06} they are small. It would be desirable to have additional experimental data for the elastic constants near the density $\rho=19.36$ \AA$^{-3}$ to confirm our finds.

Our results show that anomalies of the elastic constants are properties of a defect-free single crystal of $^4$He in an \textsl{hcp} structure.  We stress that the calculations have not considered $^3$He atoms impurities or the presence of defects. The relations among the elastic constants we have used present deviations consistent with the behavior of the supersolid fraction as a function of the system density, an increase of its value up to near $\rho = 0.0311$ \AA$^{-3}$ followed by a decrease in its value. These facts corroborate that supersolidity and shear anomalies are related phenomena probably a manifestation of a single physical process.

Comprehension of elastic properties of solids are important for  materials engineering and for basic science as well. Elasticity is one of the next chapters where the investigation of solid $^4$He will bring important contributions to quantum many-body and condensed matter physics.

We thank ﻿Norbert Mulders for comments on an earlier draft of the manuscript and valuable suggestions.The authors acknowledge financial support from the Brazilian agencies FAPESP, CNPq and CAPES. Part of the computations were performed at
the CENAPAD high-performance computing facility at Universidade Estadual
de Campinas and at the laboratory of scientific computation of Universidade Federal de Goi\'as.


\begin{thebibliography}{10}%
\makeatletter
\providecommand \@ifxundefined [1]{%
 \ifx #1\undefined \expandafter \@firstoftwo
 \else \expandafter \@secondoftwo
\fi
}%
\providecommand \@ifnum [1]{%
 \ifnum #1\expandafter \@firstoftwo
 \else \expandafter \@secondoftwo
\fi
}%
\providecommand \enquote [1]{``#1''}%
\providecommand \bibnamefont  [1]{#1}%
\providecommand \bibfnamefont [1]{#1}%
\providecommand \citenamefont [1]{#1}%
\providecommand\href[0]{\@sanitize\@href}%
\providecommand\@href[1]{\endgroup\@@startlink{#1}\endgroup\@@href}%
\providecommand\@@href[1]{#1\@@endlink}%
\providecommand \@sanitize [0]{\begingroup\catcode`\&12\catcode`\#12\relax}%
\@ifxundefined \pdfoutput {\@firstoftwo}{%
 \@ifnum{\z@=\pdfoutput}{\@firstoftwo}{\@secondoftwo}%
}{%
 \providecommand\@@startlink[1]{\leavevmode}%
 \providecommand\@@endlink[0]{}%
}{%
 \providecommand\@@startlink[1]{%
  \leavevmode
  \pdfstartlink
   attr{/Border[0 0 1 ]/H/I/C[0 1 1]}%
   user{/Subtype/Link/A<</Type/Action/S/URI/URI(#1)>>}%
  \relax
 }%
 \providecommand\@@endlink[0]{\pdfendlink}%
}%
\providecommand \url  [0]{\begingroup\@sanitize \@url }%
\providecommand \@url [1]{\endgroup\@href {#1}{\urlprefix}}%
\providecommand \urlprefix [0]{URL }%
\providecommand \Eprint[0]{\href }%
\@ifxundefined \urlstyle {%
  \providecommand \doi [1]{doi:\discretionary{}{}{}#1}%
}{%
  \providecommand \doi [0]{doi:\discretionary{}{}{}\begingroup
  \urlstyle{rm}\Url }%
}%
\providecommand \doibase [0]{http://dx.doi.org/}%
\providecommand \Doi[1]{\href{\doibase#1}}%
\providecommand \bibAnnote [3]{%
  \BibitemShut{#1}%
  \begin{quotation}\noindent
    \textsc{Key:}\ #2\\\textsc{Annotation:}\ #3%
  \end{quotation}%
}%
\providecommand \bibAnnoteFile [2]{%
  \IfFileExists{#2}{\bibAnnote {#1} {#2} {\input{#2}}}{}%
}%
\providecommand \typeout [0]{\immediate \write \m@ne }%
\providecommand \selectlanguage [0]{\@gobble}%
\providecommand \bibinfo [0]{\@secondoftwo}%
\providecommand \bibfield [0]{\@secondoftwo}%
\providecommand \translation [1]{[#1]}%
\providecommand \BibitemOpen[0]{}%
\providecommand \bibitemStop [0]{}%
\providecommand \bibitemNoStop [0]{.\EOS\space}%
\providecommand \EOS [0]{\spacefactor3000\relax}%
\providecommand \BibitemShut [1]{\csname bibitem#1\endcsname}%
\bibitem{day07}%
  \BibitemOpen
  \bibfield{author}{%
  \bibinfo {author} {\bibfnamefont{J.}~\bibnamefont{Day}}\ and\ \bibinfo
  {author} {\bibfnamefont{J.}~\bibnamefont{Beamish}},\ }%
  \bibfield{journal}{%
  \Doi{10.1038/nature06383}{\bibinfo {journal} {Nature}}\ }%
  \textbf{\bibinfo {volume} {450}},\ \bibinfo {pages} {853} (\bibinfo {month}
  {DEC 6}\ \bibinfo {year} {2007}),\ ISSN \bibinfo {issn} {{0028-0836}}%
  \bibAnnoteFile{NoStop}{day07}%
\bibitem{kim04}%
  \BibitemOpen
  \bibfield{author}{%
  \bibinfo {author} {\bibfnamefont{E.}~\bibnamefont{Kim}}\ and\ \bibinfo
  {author} {\bibfnamefont{M.~H.~W.}\ \bibnamefont{Chan}},\ }%
  \bibfield{journal}{%
  \bibinfo {journal} {Nature}\ }%
  \textbf{\bibinfo {volume} {427}},\ \bibinfo {pages} {225} (\bibinfo {month}
  {Jan.}\ \bibinfo {year} {2004}),\ ISSN \bibinfo {issn} {0028-0836},\
  \url{http://dx.doi.org/10.1038/nature02220}%
  \bibAnnoteFile{NoStop}{kim04}%
\bibitem{kim04b}%
  \BibitemOpen
  \bibfield{author}{%
  \bibinfo {author} {\bibfnamefont{E.}~\bibnamefont{Kim}}\ and\ \bibinfo
  {author} {\bibfnamefont{M.~H.~W.}\ \bibnamefont{Chan}},\ }%
  \bibfield{journal}{%
  \bibinfo {journal} {Science}\ }%
  \textbf{\bibinfo {volume} {305}},\ \bibinfo {pages} {1941} (\bibinfo {year}
  {2004}),\
  \Eprint{http://arxiv.org/abs/http://www.sciencemag.org/cgi/reprint/305/5692/%
1941.pdf}{http://www.sciencemag.org/cgi/reprint/305/5692/1941.pdf},\
  \url{http://www.sciencemag.org/cgi/content/abstract/305/5692/1941}%
  \bibAnnoteFile{NoStop}{kim04b}%
\bibitem{kim06}%
  \BibitemOpen
  \bibfield{author}{%
  \bibinfo {author} {\bibfnamefont{E.}~\bibnamefont{Kim}}\ and\ \bibinfo
  {author} {\bibfnamefont{M.~H.~W.}\ \bibnamefont{Chan}},\ }%
  \bibfield{journal}{%
  \bibinfo {journal} {Phys. Rev. Lett.}\ }%
  \textbf{\bibinfo {volume} {97}},\ \bibinfo {eid} {115302} (\bibinfo {year}
  {2006}),\ \url{http://link.aps.org/abstract/PRL/v97/e115302}%
  \bibAnnoteFile{NoStop}{kim06}%
\bibitem{day10}%
  \BibitemOpen
  \bibfield{author}{%
  \bibinfo {author} {\bibfnamefont{J.}~\bibnamefont{Day}}, \bibinfo {author}
  {\bibfnamefont{O.}~\bibnamefont{Syshchenko}},\ and\ \bibinfo {author}
  {\bibfnamefont{J.}~\bibnamefont{Beamish}},\ }%
  \bibfield{journal}{%
  \Doi{10.1103/PhysRevLett.104.075302}{\bibinfo {journal} {Phys. Rev. Lett.}}\
  }%
  \textbf{\bibinfo {volume} {104}},\ \bibinfo {pages} {075302} (\bibinfo
  {month} {Feb}\ \bibinfo {year} {2010})%
  \bibAnnoteFile{NoStop}{day10}%
\bibitem{caz12}%
  \BibitemOpen
  \bibfield{author}{%
  \bibinfo {author} {\bibfnamefont{C.}~\bibnamefont{Cazorla}}, \bibinfo
  {author} {\bibfnamefont{Y.}~\bibnamefont{Lutsyshyn}},\ and\ \bibinfo {author}
  {\bibfnamefont{J.}~\bibnamefont{Boronat}},\ }%
  \bibfield{journal}{%
  \Doi{10.1103/PhysRevB.85.024101}{\bibinfo {journal} {Phys. Rev. B}}\ }%
  \textbf{\bibinfo {volume} {85}},\ \bibinfo {pages} {024101} (\bibinfo {month}
  {Jan}\ \bibinfo {year} {2012}),\
  \url{http://link.aps.org/doi/10.1103/PhysRevB.85.024101}%
  \bibAnnoteFile{NoStop}{caz12}%
\bibitem{ros09}%
  \BibitemOpen
  \bibfield{author}{%
  \bibinfo {author} {\bibfnamefont{M.}~\bibnamefont{Rossi}}, \bibinfo {author}
  {\bibfnamefont{M.}~\bibnamefont{Nava}}, \bibinfo {author}
  {\bibfnamefont{L.}~\bibnamefont{Reatto}},\ and\ \bibinfo {author}
  {\bibfnamefont{D.~E.}\ \bibnamefont{Galli}},\ }%
  \bibfield{journal}{%
  \Doi{10.1063/1.3247833}{\bibinfo {journal} {The Journal of Chemical
  Physics}}\ }%
  \textbf{\bibinfo {volume} {131}},\ \bibinfo {eid} {154108} (\bibinfo {year}
  {2009}),\ \url{http://link.aip.org/link/?JCP/131/154108/1}%
  \bibAnnoteFile{NoStop}{ros09}%
\bibitem{uje05}%
  \BibitemOpen
  \bibfield{author}{%
  \bibinfo {author} {\bibfnamefont{S.}~\bibnamefont{Ujevic}}\ and\ \bibinfo
  {author} {\bibfnamefont{S.~A.}\ \bibnamefont{Vitiello}},\ }%
  \bibfield{journal}{%
  \bibinfo {journal} {Physical Review B (Condensed Matter and Materials
  Physics)}\ }%
  \textbf{\bibinfo {volume} {71}},\ \bibinfo {eid} {224518} (\bibinfo {year}
  {2005}),\ \url{http://link.aps.org/abstract/PRB/v71/e224518}%
  \bibAnnoteFile{NoStop}{uje05}%
\bibitem{uje06}%
  \BibitemOpen
  \bibfield{author}{%
  \bibinfo {author} {\bibfnamefont{S.}~\bibnamefont{Ujevic}}\ and\ \bibinfo
  {author} {\bibfnamefont{S.~A.}\ \bibnamefont{Vitiello}},\ }%
  \bibfield{journal}{%
  \bibinfo {journal} {Physical Review B (Condensed Matter and Materials
  Physics)}\ }%
  \textbf{\bibinfo {volume} {73}},\ \bibinfo {eid} {012511} (\bibinfo {year}
  {2006}),\ \url{http://link.aps.org/abstract/PRB/v73/e012511}%
  \bibAnnoteFile{NoStop}{uje06}%
\bibitem{azi79}%
  \BibitemOpen
  \bibfield{author}{%
  \bibinfo {author} {\bibfnamefont{R.~A.}\ \bibnamefont{Aziz}}, \bibinfo
  {author} {\bibfnamefont{V.~P.~S.}\ \bibnamefont{Nain}}, \bibinfo {author}
  {\bibfnamefont{J.~S.}\ \bibnamefont{Carley}}, \bibinfo {author}
  {\bibfnamefont{W.~L.}\ \bibnamefont{Taylor}},\ and\ \bibinfo {author}
  {\bibfnamefont{G.~T.}\ \bibnamefont{McConville}},\ }%
  \bibfield{journal}{%
  \Doi{10.1063/1.438007}{\bibinfo {journal} {J. Chem. Phys.}}\ }%
  \textbf{\bibinfo {volume} {70}},\ \bibinfo {pages} {4330} (\bibinfo {year}
  {1979}),\ \url{http://link.aip.org/link/?JCP/70/4330/1}%
  \bibAnnoteFile{NoStop}{azi79}%
\bibitem{del11}%
  \BibitemOpen
  \bibfield{author}{%
  \bibinfo {author} {\bibfnamefont{A.}~\bibnamefont{Del~Maestro}}, \bibinfo
  {author} {\bibfnamefont{M.}~\bibnamefont{Boninsegni}},\ and\ \bibinfo
  {author} {\bibfnamefont{I.}~\bibnamefont{Affleck}},\ }%
  \bibfield{journal}{%
  \Doi{10.1103/PhysRevLett.106.105303}{\bibinfo {journal} {Phys. Rev. Lett.}}\
  }%
  \textbf{\bibinfo {volume} {106}},\ \bibinfo {pages} {105303} (\bibinfo
  {month} {Mar}\ \bibinfo {year} {2011})%
  \bibAnnoteFile{NoStop}{del11}%
\bibitem{pes10}%
  \BibitemOpen
  \bibfield{author}{%
  \bibinfo {author} {\bibfnamefont{R.}~\bibnamefont{Pessoa}}, \bibinfo {author}
  {\bibfnamefont{S.~A.}\ \bibnamefont{Vitiello}},\ and\ \bibinfo {author}
  {\bibfnamefont{M.}~\bibnamefont{de~Koning}},\ }%
  \bibfield{journal}{%
  \Doi{10.1103/PhysRevLett.104.085301}{\bibinfo {journal} {Phys. Rev. Lett.}}\
  }%
  \textbf{\bibinfo {volume} {104}},\ \bibinfo {pages} {085301} (\bibinfo
  {month} {Feb}\ \bibinfo {year} {2010}),\
  \url{http://link.aps.org/doi/10.1103/PhysRevLett.104.085301}%
  \bibAnnoteFile{NoStop}{pes10}%
\bibitem{hur07}%
  \BibitemOpen
  \bibfield{author}{%
  \bibinfo {author} {\bibfnamefont{J.~J.}\ \bibnamefont{Hurly}}\ and\ \bibinfo
  {author} {\bibfnamefont{J.~B.}\ \bibnamefont{Mehl}},\ }%
  \bibfield{journal}{%
  \bibinfo {journal} {Journal of Research of the National Institute of
  Standards and Technology}\ }%
  \textbf{\bibinfo {volume} {112}},\ \bibinfo {pages} {75} (\bibinfo {month}
  {March-April 2007}\ \bibinfo {year} {2007}),\
  \url{http://nvl.nist.gov/pub/nistpubs/jres/112/2/cnt112-2.htm}%
  \bibAnnoteFile{NoStop}{hur07}%
\bibitem{jan97}%
  \BibitemOpen
  \bibfield{author}{%
  \bibinfo {author} {\bibfnamefont{A.~R.}\ \bibnamefont{Janzen}}\ and\ \bibinfo
  {author} {\bibfnamefont{R.~A.}\ \bibnamefont{Aziz}},\ }%
  \bibfield{journal}{%
  \Doi{10.1063/1.474444}{\bibinfo {journal} {The Journal of Chemical Physics}}\
  }%
  \textbf{\bibinfo {volume} {107}},\ \bibinfo {pages} {914} (\bibinfo {year}
  {1997}),\ \url{http://link.aip.org/link/?JCP/107/914/1}%
  \bibAnnoteFile{NoStop}{jan97}%
\bibitem{vit88}%
  \BibitemOpen
  \bibfield{author}{%
  \bibinfo {author} {\bibfnamefont{S.}~\bibnamefont{Vitiello}}, \bibinfo
  {author} {\bibfnamefont{K.}~\bibnamefont{Runge}},\ and\ \bibinfo {author}
  {\bibfnamefont{M.~H.}\ \bibnamefont{Kalos}},\ }%
  \bibfield{journal}{%
  \Doi{10.1103/PhysRevLett.60.1970}{\bibinfo {journal} {Phys. Rev. Lett.}}\ }%
  \textbf{\bibinfo {volume} {60}},\ \bibinfo {pages} {1970} (\bibinfo {year}
  {1988}),\ \url{http://link.aps.org/abstract/PRL/v60/p1970}%
  \bibAnnoteFile{NoStop}{vit88}%
\bibitem{mac94}%
  \BibitemOpen
  \bibfield{author}{%
  \bibinfo {author} {\bibfnamefont{T.}~\bibnamefont{MacFarland}}, \bibinfo
  {author} {\bibfnamefont{S.~A.}\ \bibnamefont{Vitiello}}, \bibinfo {author}
  {\bibfnamefont{L.}~\bibnamefont{Reatto}}, \bibinfo {author}
  {\bibfnamefont{G.~V.}\ \bibnamefont{Chester}},\ and\ \bibinfo {author}
  {\bibfnamefont{M.~H.}\ \bibnamefont{Kalos}},\ }%
  \bibfield{journal}{%
  \Doi{10.1103/PhysRevB.50.13577}{\bibinfo {journal} {Phys. Rev. B}}\ }%
  \textbf{\bibinfo {volume} {50}},\ \bibinfo {pages} {13577} (\bibinfo {month}
  {Nov}\ \bibinfo {year} {1994})%
  \bibAnnoteFile{NoStop}{mac94}%
\bibitem{pes09br}%
  \BibitemOpen
  \bibfield{author}{%
  \bibinfo {author} {\bibfnamefont{R.}~\bibnamefont{Pessoa}}, \bibinfo {author}
  {\bibfnamefont{M.}~\bibnamefont{de~Koning}},\ and\ \bibinfo {author}
  {\bibfnamefont{S.~A.}\ \bibnamefont{Vitiello}},\ }%
  \bibfield{journal}{%
  \Doi{10.1103/PhysRevB.80.172302}{\bibinfo {journal} {Physical Review B
  (Condensed Matter and Materials Physics)}}\ }%
  \textbf{\bibinfo {volume} {80}},\ \bibinfo {eid} {172302} (\bibinfo {year}
  {2009}),\ \url{http://link.aps.org/abstract/PRB/v80/e172302}%
  \bibAnnoteFile{NoStop}{pes09br}%
\bibitem{ros08}%
  \BibitemOpen
  \bibfield{author}{%
  \bibinfo {author} {\bibfnamefont{M.}~\bibnamefont{Rossi}}, \bibinfo {author}
  {\bibfnamefont{E.}~\bibnamefont{Vitali}}, \bibinfo {author}
  {\bibfnamefont{D.}~\bibnamefont{Galli}},\ and\ \bibinfo {author}
  {\bibfnamefont{L.}~\bibnamefont{Reatto}},\ }%
  \bibfield{journal}{%
  \bibinfo {journal} {J. Low Temp. Phys.}\ }%
  \textbf{\bibinfo {volume} {153}},\ \bibinfo {pages} {250} (\bibinfo {month}
  {Dec.}\ \bibinfo {year} {2008}),\
  \url{http://dx.doi.org/10.1007/s10909-008-9830-6}%
  \bibAnnoteFile{NoStop}{ros08}%
\bibitem{par80}%
  \BibitemOpen
  \bibfield{author}{%
  \bibinfo {author} {\bibfnamefont{M.}~\bibnamefont{Parrinello}}\ and\ \bibinfo
  {author} {\bibfnamefont{A.}~\bibnamefont{Rahman}},\ }%
  \bibfield{journal}{%
  \Doi{10.1103/PhysRevLett.45.1196}{\bibinfo {journal} {Phys. Rev. Lett.}}\ }%
  \textbf{\bibinfo {volume} {45}},\ \bibinfo {pages} {1196} (\bibinfo {month}
  {Oct}\ \bibinfo {year} {1980})%
  \bibAnnoteFile{NoStop}{par80}%
\bibitem{par81}%
  \BibitemOpen
  \bibfield{author}{%
  \bibinfo {author} {\bibfnamefont{M.}~\bibnamefont{Parrinello}}\ and\ \bibinfo
  {author} {\bibfnamefont{A.}~\bibnamefont{Rahman}},\ }%
  \bibfield{journal}{%
  \Doi{10.1063/1.328693}{\bibinfo {journal} {Journal of Applied Physics}}\ }%
  \textbf{\bibinfo {volume} {52}},\ \bibinfo {pages} {7182} (\bibinfo {year}
  {1981}),\ \url{http://link.aip.org/link/?JAP/52/7182/1}%
  \bibAnnoteFile{NoStop}{par81}%
\bibitem{cep77}%
  \BibitemOpen
  \bibfield{author}{%
  \bibinfo {author} {\bibfnamefont{D.}~\bibnamefont{Ceperley}}, \bibinfo
  {author} {\bibfnamefont{G.~V.}\ \bibnamefont{Chester}},\ and\ \bibinfo
  {author} {\bibfnamefont{M.~H.}\ \bibnamefont{Kalos}},\ }%
  \bibfield{journal}{%
  \Doi{10.1103/PhysRevB.16.3081}{\bibinfo {journal} {Phys. Rev. B}}\ }%
  \textbf{\bibinfo {volume} {16}},\ \bibinfo {pages} {3081} (\bibinfo {month}
  {Oct}\ \bibinfo {year} {1977})%
  \bibAnnoteFile{NoStop}{cep77}%
\bibitem{sti01}%
  \BibitemOpen
  \bibfield{author}{%
  \bibinfo {author} {\bibfnamefont{L.}~\bibnamefont{Stixrude}},\ }%
  \enquote{\bibinfo {title} {Handbook of elastic properties of solids, liquids,
  and gases},}\ \ (\bibinfo {publisher} {Academic Press},\ \bibinfo {address}
  {London},\ \bibinfo {year} {2001})\ Chap.\ \bibinfo {chapter} {Elasticity of
  oxides and ionics}, pp.\ \bibinfo {pages} {31--56}%
  \bibAnnoteFile{NoStop}{sti01}%
\bibitem{fra70}%
  \BibitemOpen
  \bibfield{author}{%
  \bibinfo {author} {\bibfnamefont{J.~P.}\ \bibnamefont{Franck}}\ and\ \bibinfo
  {author} {\bibfnamefont{R.}~\bibnamefont{Wanner}},\ }%
  \bibfield{journal}{%
  \Doi{10.1103/PhysRevLett.25.345}{\bibinfo {journal} {Phys. Rev. Lett.}}\ }%
  \textbf{\bibinfo {volume} {25}},\ \bibinfo {pages} {345} (\bibinfo {month}
  {Aug}\ \bibinfo {year} {1970})%
  \bibAnnoteFile{NoStop}{fra70}%
\bibitem{bor54}%
  \BibitemOpen
  \bibfield{author}{%
  \bibinfo {author} {\bibfnamefont{M.}~\bibnamefont{Born}}\ and\ \bibinfo
  {author} {\bibfnamefont{K.}~\bibnamefont{Huang}},\ }%
  \emph{\bibinfo {title} {Dynamical Theory of Crystal Lattices}},\
  International Series of Monographs on Physics\ (\bibinfo {publisher} {Oxford
  University Press},\ \bibinfo {address} {Walton Street, Oxford OX2 6DP, UK},\
  \bibinfo {year} {1954})\ pp.\ \bibinfo {pages} {xii + 420}%
  \bibAnnoteFile{NoStop}{bor54}%
\bibitem{nye85}%
  \BibitemOpen
  \bibfield{author}{%
  \bibinfo {author} {\bibfnamefont{J.}~\bibnamefont{Nye}},\ }%
  \emph{\bibinfo {title} {Physical properties of crystals: their representation
  by tensors and matrices}}\ (\bibinfo {publisher} {Oxford University Press},\
  \bibinfo {address} {Walton Street, Oxford OX2 6DP, UK},\ \bibinfo {year}
  {1985})%
  \bibAnnoteFile{NoStop}{nye85}%
\bibitem{zha04}%
  \BibitemOpen
  \bibfield{author}{%
  \bibinfo {author} {\bibfnamefont{C.-S.}\ \bibnamefont{Zha}}, \bibinfo
  {author} {\bibfnamefont{H.-k.}\ \bibnamefont{Mao}},\ and\ \bibinfo {author}
  {\bibfnamefont{R.~J.}\ \bibnamefont{Hemley}},\ }%
  \bibfield{journal}{%
  \Doi{10.1103/PhysRevB.70.174107}{\bibinfo {journal} {Phys. Rev. B}}\ }%
  \textbf{\bibinfo {volume} {70}},\ \bibinfo {pages} {174107} (\bibinfo {month}
  {Nov}\ \bibinfo {year} {2004})%
  \bibAnnoteFile{NoStop}{zha04}%
\bibitem{cre71}%
  \BibitemOpen
  \bibfield{author}{%
  \bibinfo {author} {\bibfnamefont{R.~H.}\ \bibnamefont{Crepeau}}, \bibinfo
  {author} {\bibfnamefont{O.}~\bibnamefont{Heybey}}, \bibinfo {author}
  {\bibfnamefont{D.~M.}\ \bibnamefont{Lee}},\ and\ \bibinfo {author}
  {\bibfnamefont{S.~A.}\ \bibnamefont{Strauss}},\ }%
  \bibfield{journal}{%
  \Doi{10.1103/PhysRevA.3.1162}{\bibinfo {journal} {Phys. Rev. A}}\ }%
  \textbf{\bibinfo {volume} {3}},\ \bibinfo {pages} {1162} (\bibinfo {month}
  {Mar}\ \bibinfo {year} {1971})%
  \bibAnnoteFile{NoStop}{cre71}%
\bibitem{gre71}%
  \BibitemOpen
  \bibfield{author}{%
  \bibinfo {author} {\bibfnamefont{D.~S.}\ \bibnamefont{Greywall}},\ }%
  \bibfield{journal}{%
  \Doi{10.1103/PhysRevA.3.2106}{\bibinfo {journal} {Phys. Rev. A}}\ }%
  \textbf{\bibinfo {volume} {3}},\ \bibinfo {pages} {2106} (\bibinfo {month}
  {Jun}\ \bibinfo {year} {1971})%
  \bibAnnoteFile{NoStop}{gre71}%
\bibitem{wan70}%
  \BibitemOpen
  \bibfield{author}{%
  \bibinfo {author} {\bibfnamefont{R.}~\bibnamefont{Wanner}}\ and\ \bibinfo
  {author} {\bibfnamefont{J.~P.}\ \bibnamefont{Franck}},\ }%
  \bibfield{journal}{%
  \Doi{10.1103/PhysRevLett.24.365}{\bibinfo {journal} {Phys. Rev. Lett.}}\ }%
  \textbf{\bibinfo {volume} {24}},\ \bibinfo {pages} {365} (\bibinfo {month}
  {Feb}\ \bibinfo {year} {1970})%
  \bibAnnoteFile{NoStop}{wan70}%
\bibitem{ree71}%
  \BibitemOpen
  \bibfield{author}{%
  \bibinfo {author} {\bibfnamefont{R.~A.}\ \bibnamefont{Reese}}, \bibinfo
  {author} {\bibfnamefont{S.~K.}\ \bibnamefont{Sinha}}, \bibinfo {author}
  {\bibfnamefont{T.~O.}\ \bibnamefont{Brun}},\ and\ \bibinfo {author}
  {\bibfnamefont{C.~R.}\ \bibnamefont{Tilford}},\ }%
  \bibfield{journal}{%
  \Doi{10.1103/PhysRevA.3.1688}{\bibinfo {journal} {Phys. Rev. A}}\ }%
  \textbf{\bibinfo {volume} {3}},\ \bibinfo {pages} {1688} (\bibinfo {month}
  {May}\ \bibinfo {year} {1971})%
  \bibAnnoteFile{NoStop}{ree71}%
\bibitem{bea01}%
  \BibitemOpen
  \bibfield{author}{%
  \bibinfo {author} {\bibfnamefont{J.~R.}\ \bibnamefont{Beamish}},\ }%
  \enquote{\bibinfo {title} {Handbook of elastic properties of solids, liquids
  and gases},}\ \ (\bibinfo {publisher} {Academic Press},\ \bibinfo {address}
  {London},\ \bibinfo {year} {2001})\ Chap.\ \bibinfo {chapter} {Solid Inert
  Gases}, pp.\ \bibinfo {pages} {77--95}%
  \bibAnnoteFile{NoStop}{bea01}%
\bibitem{edw65}%
  \BibitemOpen
  \bibfield{author}{%
  \bibinfo {author} {\bibfnamefont{D.~O.}\ \bibnamefont{Edwards}}\ and\
  \bibinfo {author} {\bibfnamefont{R.~C.}\ \bibnamefont{Pandorf}},\ }%
  \bibfield{journal}{%
  \Doi{10.1103/PhysRev.140.A816}{\bibinfo {journal} {Phys. Rev.}}\ }%
  \textbf{\bibinfo {volume} {140}},\ \bibinfo {pages} {A816} (\bibinfo {month}
  {Nov}\ \bibinfo {year} {1965})%
  \bibAnnoteFile{NoStop}{edw65}%
\bibitem{lou93}%
  \BibitemOpen
  \bibfield{author}{%
  \bibinfo {author} {\bibfnamefont{P.}~\bibnamefont{Loubeyre}}, \bibinfo
  {author} {\bibfnamefont{R.}~\bibnamefont{LeToullec}}, \bibinfo {author}
  {\bibfnamefont{J.~P.}\ \bibnamefont{Pinceaux}}, \bibinfo {author}
  {\bibfnamefont{H.~K.}\ \bibnamefont{Mao}}, \bibinfo {author}
  {\bibfnamefont{J.}~\bibnamefont{Hu}},\ and\ \bibinfo {author}
  {\bibfnamefont{R.~J.}\ \bibnamefont{Hemley}},\ }%
  \bibfield{journal}{%
  \bibinfo {journal} {Phys.\ Rev.\ Lett.}\ }%
  \textbf{\bibinfo {volume} {71}},\ \bibinfo {pages} {2272} (\bibinfo {year}
  {1993})%
  \bibAnnoteFile{NoStop}{lou93}%
\bibitem{uje03}%
  \BibitemOpen
  \bibfield{author}{%
  \bibinfo {author} {\bibfnamefont{S.}~\bibnamefont{Ujevic}}\ and\ \bibinfo
  {author} {\bibfnamefont{S.~A.}\ \bibnamefont{Vitiello}},\ }%
  \bibfield{journal}{%
  \bibinfo {journal} {J. Chem. Phys.}\ }%
  \textbf{\bibinfo {volume} {119}},\ \bibinfo {pages} {8482} (\bibinfo {year}
  {2003}),\ \url{http://link.aip.org/link/?JCP/119/8482/1}%
  \bibAnnoteFile{NoStop}{uje03}%
\end{thebibliography}

%

\end{document}